 \documentclass[final,1p,times,twocolumn]{elsarticle}

\usepackage{subfigure}

\usepackage{amssymb}
\usepackage{aas_macros}

\biboptions{square, comma}

\journal{Astroparticle Physics}

\begin{document}

\begin{frontmatter}



\title{A Simple Method to Test for Energy-Dependent Dispersion in High Energy 
Light-Curves of Astrophysical Sources}


\author[durham]{Ulisses Barres de Almeida\fnref{1}}
\ead{ulisses@mppmu.mpg.de}
\fntext[1]{Now at Max-Planck-Institut f\"{u}r Physik, D-80805, M\"{u}nchen, Deutschland.}
\author[durham]{M.~ K.~Daniel\corref{1}}
\ead{michael.daniel@durham.ac.uk}
\cortext[1]{+44 191 334 3678}

\address[durham]{Department of Physics, University of Durham, South Road, Durham, DH1 3LE, England}

\begin{abstract}
In this paper we discuss a simple method of testing for the presence of 
energy-dependent dispersion in high energy data-sets. It uses the minimisation 
of the Kolmogorov distance between the cumulative distribution of two 
probability functions as the statistical metric to estimate the magnitude of any
spectral dispersion within transient features in a light-curve and we also show 
that it performs well in the presence of modest energy resolutions ($\sim 20\%$) 
typical of gamma-ray observations. After presenting the method in detail we 
apply it to a parameterised simulated lightcurve based on the extreme VHE 
gamma-ray flare of PKS\,2155-304 observed with H.E.S.S. in 2006, in order to 
illustrate its potential through the concrete example of setting constraints on 
quantum-gravity induced Lorentz invariance violation (LIV) effects. We obtain 
comparable limits to those of the most advanced techniques used in LIV searches 
applied to similar datasets, but the present method has the advantage of being 
particularly straightforward to use. Whilst the development of the method was 
motivated by LIV searches, it is also applicable to other astrophysical 
situations where energy-dependent dispersion is expected, such as spectral lags 
from the acceleration and cooling of particles in relativistic outflows.
\end{abstract}


\end{frontmatter}


\section{Introduction}
\label{sec:intro}
The timing properties of a sequence of events can be very revealing as to the 
physical nature either of the emitting source or of the medium they propagate 
through, especially when taken in conjunction with information about their
energy. 
Timing analysis algorithms with the capacity of resolving energy-dependent 
properties can then be an important tool for probing the physical mechanisms 
leading to flux variability, such as particle acceleration and cooling. 
Methods are traditionally based on cross-correlation of the binned time-series 
(e.g. \cite{xcol2155}), and sometimes rely on a particular parameterisation of 
the light-curve, for example by modeling the data according to a pre-determined 
choice for the light-curve profile (e.g. \cite{likelihood2155}). In the case of 
gamma-ray sources, where high-energy processes are responsible for extreme and 
short-lived variability events, and for which the observational data are often
limited by low photon statistics, unbinned methods are the natural and 
preferential choice of approach to the problem of temporal analysis of time- and
energy-stamped photon lists.

In this paper we present a method to search for energy-dependent 
dispersion in light-curves of transient sources with relatively sparse events, 
particularly suited for (though of course not limited to) ground-based 
gamma-ray telescopes. A particular motivation for the study of energy dependent 
dispersion in the very-high energy (VHE) regime is the prospect of testing 
for possible signatures of Lorentz invariance violation (LIV), foreseen by a 
number of theories of quantum gravity (QG) \cite{Sarkar02}. The analysis method 
is described in section~\ref{sec:UnbinnedTests} and its suitability for the 
particularly challenging task of searching for LIV effects are illustrated in 
section~\ref{sec:performance} and applied to AGN lightcurves in 
section~\ref{sec:PKS2155}.

\subsection{Lorentz invariance violation}
\label{sec:LIV}
The unification of the theories of quantum mechanics, governing the smallest of 
scales, and that of gravity, governing the largest of scales, is one of the most
serious challenges in modern physics. Because of the extremely high energies at 
which QG effects are expected to manifest (around the Planck scale, 
$E_{\rm{QG}} \approx E_{\rm{P}} \simeq 10^{19}$ GeV) the effects are only likely
to become noticeable at very high energies and so difficult to be assessed 
directly in the laboratory. The so-called time-of-flight experiments, first 
proposed in a seminal paper by Amelino-Camelia et al. (1998) 
\cite{Amelino-Camelia98}, is one of the most promising ways of carrying out 
tests for QG signatures. The method is based on the search for an 
energy-dependent speed of light ($c$) in vacuum\footnote{This is because in QG 
theories the vacuum is expected to have a non-trivial refractive index due to 
fluctuations of the space-time at the quantum level.} from the observation of 
GeV-TeV photons propagating over cosmological distances. The exact form of the 
energy-dependent photon momentum due to QG effects can vary depending on the 
particular theory adopted, but given that its effect is very small it can be 
treated perturbatively leading to a form (eg, following the scheme of 
\cite{xcol2155, likelihood2155, magicqg, FermiLimits})
\begin{equation}
\label{eq:perturbation}
 c^2 p^2 = E_{\gamma}^2[1+\xi_{1} E_{\gamma}/E_{QG}
         + \xi_{2}(E_{\gamma}^2/E_{\rm{QG}}^2) + \ldots]
\end{equation}
The consequently small magnitude of its signature at astrophysically 
accessible energy ranges\footnote{The most energetic photons recorded from 
astrophysical sources have energies of $\sim$ tens of TeV and for 
$E_{\gamma} \sim 1$ TeV the correction to the speed of light due to quantum 
gravity would be of order $ 10^{-15}c$} mean that these searches require 
extremely sensitive measurements. In time-of-flight experiments the cumulative 
temporal effects of small variations in $c$ is amplified, eventually manifesting
as measurable time-delays over the integrated distance travelled by the photons.

To first order, the magnitude of the time delays expected from QG variations of 
$c$ are $\delta t \propto E_{\gamma}/E_{\rm{QG}} \sim  10$\,s/TeV/Gpc for Planck
scale QG. This implies that searches from distant sources are preferred (which
in turn can lead to them being correspondingly fainter than nearby sources) and 
that the searches should be conducted over narrow features (see 
section~\ref{sec:width} for further details). For instance, in the case of the 
active galactic nucleus PKS\,2155-304 located at a redshift $z\sim0.116$, for 
which we would expect a delay of $\delta t \sim 4$ s per TeV in photon energy, 
we would need flare features on timescales of no more than tens to hundreds of 
seconds in the VHE light-curve to bring the effect to the fore. With event rates
of a few Hz during the brightest flares \cite{bigflare} the latter property 
disfavours binning methods on count-rate limited datasets. Sensitivity to small 
spectral dispersions within very limited photon lists is therefore the most 
desired characteristic of a dispersion-search method used for time-of-flight 
measurements.

\section{Unbinned Methods - Dispersion Cancellation Algorithm}
\label{sec:UnbinnedTests}
Unbinned algorithms are well-suited for the identification and analysis of local
and aperiodic light-curve features, such as bursts or flares in AGN or GRB data.
Indeed, the observation of GeV photons from GRB~080916C and GRB~090510 by the 
Fermi/LAT collaboration \cite{GRB080916C, FermiLimits} has recently been able to
set limits at and just above the Planck scale for linear-term effects using two 
different unbinned approaches. The first of these methods was to directly 
compare the arrival time of the highest energy photon to the different burst 
features and, assuming they were contemporaneous at the source, determine what 
magnitude of dispersion would have to be experienced by the photons during 
propagation to the observer to explain the observed lag. There are two main 
drawbacks in such an approach. The first caveat comes from uncertainty in 
the knowledge of the intrinsic structure of the light-curve, due for example to
a lack of understanding of precursor activity in GRBs (see, e.g. 
\cite{FermiPrecursor}), which can cast doubt as to which particular features to 
associate with the highest energy photons upon assigning the delay. The second 
drawback has to do with the application of the method to ground-based TeV 
gamma-ray observations. The poorer energy resolution\footnote{The energy 
resolution is defined as $|\Delta E|/E$, where $\Delta E$ is the difference 
between the true energy and the analysis-reconstructed energy of an event.} of 
the ground based instruments ($|\Delta E|/E\sim15-20\%$) in comparison to the 
Fermi/LAT resolution (generally $|\Delta E|/E \lesssim 10\%$ above 1 GeV
\cite{FermiSpecs}) would mean that the uncertainty in the dispersion of a single 
photon (of a few s/TeV) could easily hide any anticipated dispersion.

A number of different approaches exist that are specifically designed for tests 
of time lags between event sequences, such as likelihood methods 
\cite{martinez09} and modified cross-correlation functions applied to the 
individual photon events \cite{MCCF}. A particularly attractive and simple 
algorithm was the second approach used in the Fermi analysis of GRB~090510. This
method was conceived to solve the problem of detecting energy-dependent time 
lags in statistically limited photon lists, and the fundamental idea of such a 
{\em dispersion cancellation} algorithm\footnote{This name was coined by 
Scargle et al. (2008) in the context of their particular version of the test, 
but we will adopt it here with greater generality} method was independently 
proposed by Scargle et al. (2008) \cite{scargle08} and Ellis et al. (2008) 
\cite{ellis08} -- the latter derived actually to search for QG signatures from 
neutrino propagation. We also use this technique as the basis for our search 
methodology, but introduce a different test metric that is better suited to the 
systematic uncertainties associated with a VHE photon dataset, in particular the
poorer energy resolution.

In general, if the expected energy-dispersion is small compared to other 
relevant variability timescales of the astrophysical system under study, its 
exact functional form is of little importance, since the dependency can be 
treated perturbatively and expressed as the first-order terms of a Taylor 
series (cf. equation \ref{eq:perturbation}). The {\em dispersion cancellation} 
algorithm uses this fact and works directly on the time- and energy-tagged 
events to search for a non-zero parameter $\tau$ (measured here in s/TeV) 
that optimally cancels any spectral dispersion present in the light-curve. The 
lag-correction, $\delta t_i$, on photon $i$ of energy $E_i$, is given by:
\begin{equation}
\label{eq:cancel}
 \delta t_i = -\tau E_i^\alpha
\end{equation}
where $\alpha$ defines the dominant term of the series expansion for the energy 
dependency of the time lag, usually taken to be the linear expansion term 
$\alpha = 1$, or the quadratic term $\alpha=2$. The {\em dispersion 
cancellation} algorithm cycles through a range of possible values for $\tau$, 
looking for the $\tau^*$ that extremises an appropriate metric (or 
``cost function''), chosen to quantify the presence of spectral lags. An 
advantage of this approach is that it makes no {\em a priori} assumptions on 
the nature of the lightcurve apart from the inevitable hypothesis of 
simultaneity of emission at the source.

\begin{figure}[htbp]
  \centering
  \includegraphics[angle=0, width=0.6\textwidth]{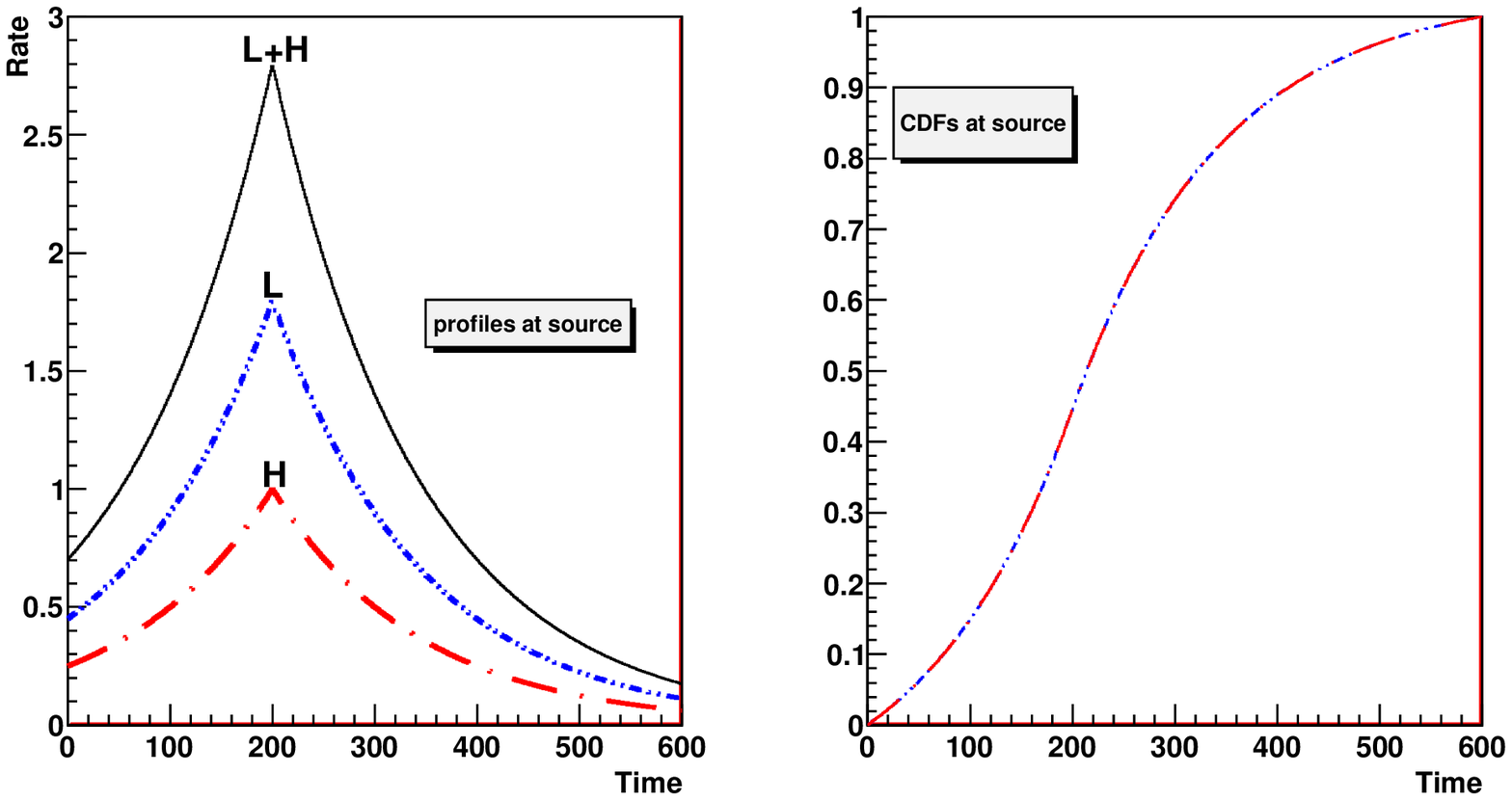}
  \includegraphics[angle=0, width=0.6\textwidth]{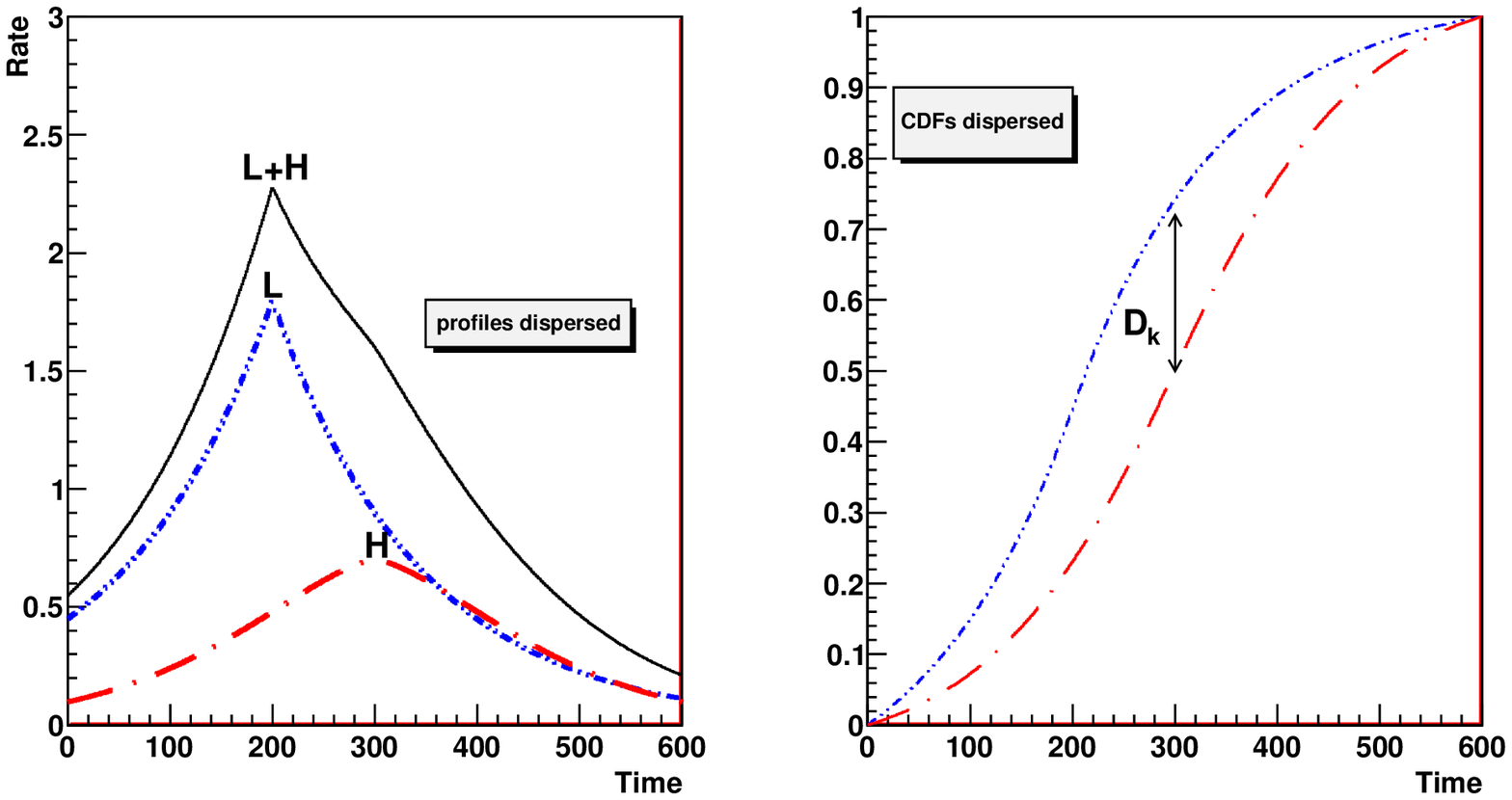}
  \caption {Cartoon of the effect of the energy dependent dispersion
    on the shape of the low (L) and high (H) energy profiles. Observe
    that the systematic shift on the high-energy curve relative to the
    low-energy one induces a skew to the burst. 
    The panels to the right show the corresponding 
    discrepancy in the cumulative distribution function. The maximum vertical 
    distance is indicated, corresponding to the Kolmogorov measure $D_{K}$. 
    Observe that $D_{K}$ tends to fall always in the middle of the distribution,
    near the peak position of the profiles.}
  \label{fig:cartoon}
\end{figure}

A number of different test metrics have been proposed for the purpose of 
quantifying the spectral lag and finding the optimal dispersion cancellation 
parameter. They all use some kind of measure of sharpness of the peak in the 
burst profile as the value to be maximised in the search for $\tau^*$ (see 
examples in \cite{ellis08}, \cite{magicqg} and \cite{scargle08}). Here by 
``sharpness'' of a burst we mean a quantity proportional to the gradient of the 
photon density at the time of the maximum in emission. The principle behind the 
maximum sharpness choice is that whilst the emission of high and 
low energy photons at the source is simultaneous (top left panel of 
figure~\ref{fig:cartoon}), an energy-dependent dispersion introduced during the 
photon propagation will always skew the overall light-curve. In the particular 
example shown in figure~\ref{fig:cartoon}, this happens by the delayed arrival 
of the higher energy photons (lower left hand plot), thus skewing and broadening 
the burst profile as a result. The maximally sharp burst configuration will be 
retrieved when the temporal sequence of events is again randomised in energy, 
corresponding to the exact cancellation of the dispersion. Observe that 
this approach will always give a unique solution for each given dispersion 
model, because in the case of under- or over-corrections, $\tau$, the 
asymmetric effect will either still be left present or be re-introduced in the 
opposite direction, and the burst will remain broadened in respect to its 
original width. 
The cost functions used in \cite{scargle08, FermiLimits} serve well to minimise 
the total inter-photon spacing within the entire event sequence, thereby 
maximising the peak of the lightcurve, but the poorer energy resolution of the 
ground-based instruments limits the efficacy of the method when applied to VHE 
observations of sharp bursts. In this paper we examine an alternative test 
metric based on the Kolmogorov distance between two probability distributions, 
which better exploits the fact that, whilst the energy resolution of an 
individual photon is far from ideal, the overall energy bias of a sample of 
them is actually $\sum_{i} \Delta E_{i} \simeq 0$.

\subsection{The Minimum Kolmogorov Distance Metric}
\label{sec:Kolmogorov}
For a data-set with a sufficient number of photons (a few tens), the event list 
can be separated into low- and high-energy bands, forming two independent 
datasets. In the absence of any spectral dispersion, the basic assumption that 
the temporal sequence of events is randomised in energy should hold and the 
profiles (apart from statistical fluctuations plus some arbitrary intensity 
scaling that can be eliminated by normalisation) should superpose. If, however, 
a systematic spectral dispersion is present, the profiles of the light-curve 
will look skewed relative to each other (see Figure~\ref{fig:cartoon}, lower 
panels). 

Given two random variables $X$ and $Y$ in $\mathbb{R}$, a simple
measure of the difference between their respective probability
distributions is the \textit{Kolmogorov distance} $D_{K}$, defined as the 
maximum vertical distance between the two cumulative distribution functions
(CDFs) \cite{kolmogorov}:
\begin{equation}
\label{eq:KD}
 D_{K} \equiv \sup_{x~\in~\mathbb{R}}|F_{X}(x) - F_{Y}(x)|
\end{equation}
where $F_{X}(x) = \rm{prob}(X \leq x)$ and $F_{Y}(x) = \rm{prob}(Y \leq x)$.

The situation is illustrated in the right-hand plots of 
figure~\ref{fig:cartoon}. Assuming that the events are generated 
simultaneously and co-spatially in the source, any energy-dependent dispersion 
introduced between the two will show up as an increased $D_{k}$ between the 
cumulative distribution functions of their arrival times. Therefore, minimising 
this value will amount to cancelling any dispersion present (simultaneously 
minimising the sharpness of the profile). It is well known from the properties 
of the Kolmogorov-Smirnov test that the Kolmogorov distance is insensitive to 
the tails of the distributions, where the CDFs converge to the values of 0 and 
1, and which describe the probability of extreme events \cite{nr}. In fact, 
$D_{K}$ will tend to fall around the central regions of the CDF, near to the 
peaks of the profiles where their accumulated discrepancy is maximum. This is a 
useful property because it means that the measure naturally attributes a greater 
weight to the most transient parts of the light-curve, whilst being relatively 
insensitive to outliers.

\section{Performance of the Method}
\label{sec:performance}
We now analyse the performance of the method by discussing the four main 
factors that are expected to affect the sensitivity to detect energy-dependent 
dispersion: burst width (section~\ref{sec:width}), energy resolution
(section~\ref{sec:Eres}), burst intensity and asymmetry 
(section~\ref{sec:asymmetry}). Before we move on to discussing these specific 
topics, we list the steps for application of the algorithm:

\begin{itemize}

\item select a burst or transient event from the light-curve; 

\item split the burst photons into low- and high-energy datasets, this will be a
trade-off such that both groups have the largest possible number of events in 
them, but also that the difference between their average energy is as large as 
possible; 

\item build the CDFs for the two distributions (see~\ref{app:lightcurve} for
some discussion on alternatives ways of representing the lightcurve); 

\item adopt a model for the time delay, e.g. linear or quadratic in energy;

\item apply correction to the time-stamp of photons according to 
equation~\ref{eq:cancel};

\item for each value $\tau$ of the correction, calculate $D_{K}$;

\item the optimum $\tau^*$ is the one which minimises $D_{K}$ for the range of 
$\tau$ tested;

\item assess the uncertainty by simulation of the burst or bootstrap of the 
data.

\end{itemize}

For illustrative purposes, we will only consider in this section the ideal case 
of an isolated Gaussian burst. The superposition of multiple bursts or burst 
shapes different from Gaussian will be discussed when the method is applied to 
real flare data from the AGN PKS\,2155-304 in the next section, but do not 
change the conclusions presented here. For our studies individual burst data 
were simulated using the generalised Gaussian shape \cite{norris96}, 
which can also provide a good match to the pulse profiles generally observed 
from AGNs and GRBs:
\begin{equation}
\label{eq:GeneralisedGaussian}
 I(t) = I_{max}\exp \left[-\left( \frac{|t-t_{max}|}{\sigma_{r,~d}}\right)^{\kappa} \right]
\end{equation}
where $t$ is the time into the flare, 
$t_{max}$ is the time of maximum flux $I_{max}$, 
$\sigma_r$ and $\sigma_d$ are the signal rise (for $t<t_{max}$) and decay 
(for $t>t_{max}$) time constants respectively, the ``sharpness'' 
(peakedness, or kurtosis) of the profile is given by the parameter $\kappa > 0$. 
A low value of $\kappa$ means a sharply peaked pulse, a high value a more 
rounded one, and $\kappa=2$ corresponds to a Gaussian pulse shape. The rise 
($t_{r}$) and decay ($t_{d}$) times from half to maximum amplitude are found 
from the rise and decay constants using the equation 
$t_{r,d} = [ln(2)^{1/\kappa}]\sigma_{r,d}$.
The spectra are assumed to take a power law shape of the form $dN/dE =
kE^{\Gamma}$, where $\Gamma$ is the spectral index and $k$ a flux constant.

\subsection{Energy cuts}
\label{sec:EnergyCuts}
The first step necessary in constructing the CDFs for the analysis means we must 
decide where to place the low- and high-energy boundaries. This choice is made 
such that the difference in the mean energy between the two CDFs can be 
maximised, while keeping good photon statistics in both energy bins for the 
analysis. We have verified that due to the (usually) steeply-falling spectral 
index of the photon distributions, the analysis is less sensitive to the choice 
of the low-energy boundary, provided this is set comfortably above the threshold 
energy of the instrument. We set here the low energy band to be
$0.2\leq\mathrm{S}\leq0.4$\,TeV. We then searched for an optimal high-energy 
cut window, which will be the more statistically-starved component. Simulations
were for a Gaussian light curve shape of 120s rise/fall time and a maximum count
rate of 3\,Hz.

\begin{figure}[htbp]
 \begin{center}
  \includegraphics[width=\textwidth]{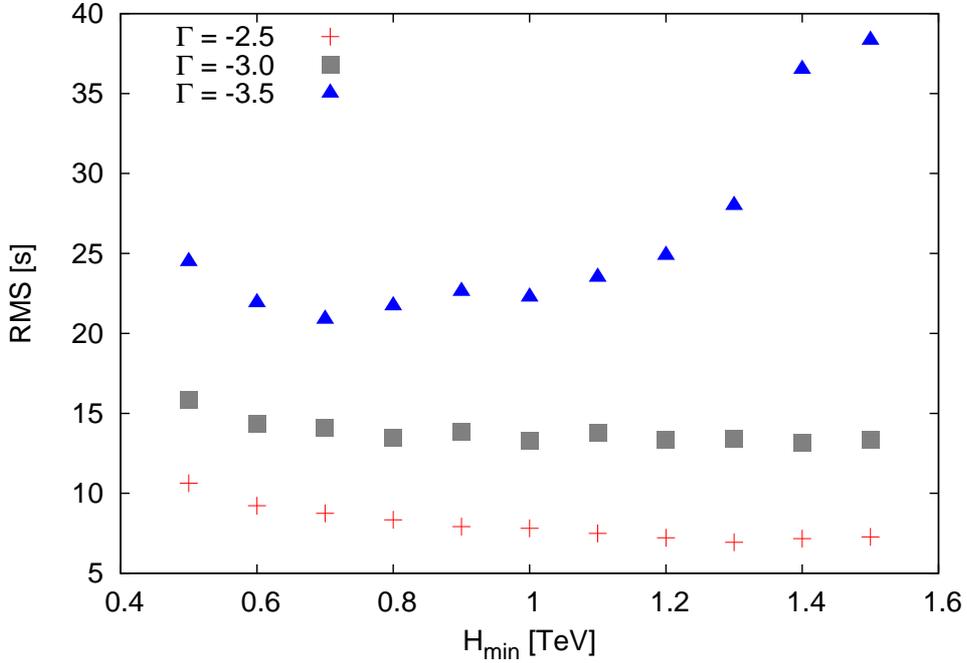}
  \caption {Effect of the choice of the minimum energy cut for the high energy 
            band ($\mathrm{H_{min}}$) on the accuracy of the determined 
            dispersion measure. Simulations are for power law spectra with
            indices of -2.5 (crosses), -3.0 (squares) and -3.5 (triangles).}
  \label{fig:EnergyCutGauss}
 \end{center}
\end{figure}

Figure~\ref{fig:EnergyCutGauss} shows the results of our analysis on the effect of 
the choice of the minimum value for the high energy cut $\mathrm{H_{min}}$ on 
the root-mean-square (RMS)\footnote{The root mean square of a distribution $X$ 
with $N$ events $x_i$  is given by $RMS = \sqrt{\sum{x_i^2}/N}$.} of the 
reconstructed dispersion parameter ($\tau^{*}$) for three different values of
the spectral index (for $\Gamma=-2.5, -3.0, -3.5$). The curves for each spectrum
show a slight improvement of the RMS with increasing energy cut which quickly 
plateaus. This is because at too low a minimum energy for the high energy sample
the difference in the average photon energy of the events gets too low and the 
high-energy CDF becomes indistinguishable from the low energy CDF as they are 
both dominated by events that are not dispersed much. The error in any 
reconstructed lag correspondingly increases as natural statistical fluctuations 
become the dominant source of uncertainty. Much more noticeable for this though,
is the steep rise in the RMS for $\mathrm{H_{min}}>1$\,TeV for the softest 
energy spectrum ($\Gamma=-3.5$). This occurs when the number of events in the 
high energy sample drops below $\sim$ 10 -- the CDF becomes ill-defined with 
respect to the low energy sample CDF and so no reliable dispersion measure can 
be found. This result gives an idea of the minimum number of events necessary 
in an energy band for the method to be able to work. From this we see that the 
results of the test will be fairly independent of the actual minimum high energy
cut, provided the high energy band has $>10$ events and is at least a factor of 
two higher in energy than the lower energy sample. We take 
$\mathrm{H_{min}} > 1$\,TeV for the remainder of the paper, unless specified
otherwise.

\subsection{Sensitivity to burst width}
\label{sec:width}
We quantify the sensitivity to the burst width by the term 
``sensitivity factor'', $\eta$, following the definition in 
\cite{Amelino-Camelia98}. This quantity is written as the ratio of the 
expected dispersion magnitude ($\delta t$) to the width of the transient feature 
($\Delta t$): 
\begin{equation}
\label{eq:sensitivity}
 \eta = \frac{\delta t}{\Delta t}
\end{equation}
This ratio is the main parameter which will quantify the size of the lag that 
can be probed by the method, for a given burst width. 

\begin{figure}[htbp]
 \begin{center}
  \includegraphics[width=0.8\textwidth]{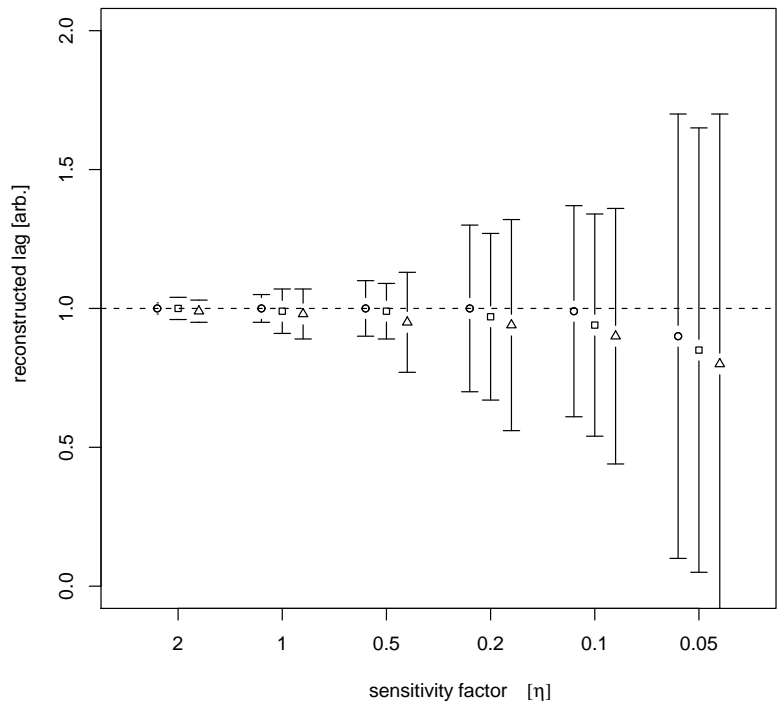}
  \caption {Sensitivity of the algorithm to the ratio lag/burst width $\eta$ for 
   0\% (open circle), 10\% (open square), and 20\% (open triangle) energy 
   resolution. The results are from sets of 10,000 Monte Carlo simulations of 
   Gaussian profiles, containing 500 events each, for a low-energy threshold of 
   0.2 TeV and spectral index $\Gamma = -2.5$. The low- and high-energy bins 
   were defined such that the average energy difference between the two is 
   $\sim 1$ TeV.}
  \label{fig:EnergyResolution}
 \end{center}
\end{figure}

For this analysis, we simulated 10,000 Gaussian burst profiles of 500 events 
each, with a low-energy threshold of 200 GeV and a spectral index 
$\Gamma = -2.5$. 
A dispersion was then introduced that varies from 5-200\% of the 
burst width, i.e. from the dispersion being entirely contained within the burst 
to the burst being smeared in time over a period greater than its duration. 
The results are shown in figure~\ref{fig:EnergyResolution}, where the points 
correspond to the mean reconstructed $\tau^{*}$ and the error bars are the RMS 
of that distribution. We see, as expected, that the narrower the width of the 
burst with respect to the introduced delay, the better the delay can be 
determined. The error bars in the plot indicate the 68\% confidence intervals 
(CI) of the reconstructed lag distribution, showing that the method can 
reconstruct the correct value of $\tau$ and exclude the null hypothesis of zero 
lag at the 99\% level up to a value of $\eta \approx 0.2$. This corresponds to 
a sensitivity limit of a lag equal to 20\% of the burst width.

\subsection{Sensitivity to energy resolution}
\label{sec:Eres}
We also included in our sensitivity analysis the effect of the energy resolution 
($|\Delta E|/E$), an important consideration in ground-based gamma-ray 
measurements. This uncertainty will directly affect the dispersion correction 
for individual photons and will thus limit the sensitivity of the method. The 
energy resolution is modelled as 0\% (an ideal detector case where the 
reconstructed energy is always the true energy), 10\% and 20\%, shown as the 
sub-sets of data in figure~\ref{fig:EnergyResolution}. There is a small 
systematic trend for the reconstructed lag to be under-estimated as the energy 
resolution gets poorer, but this is very small in comparison to the overall 
error in the reconstructed $\tau^{*}$. The under-estimation is expected: the 
power law nature of the spectrum means that any width to the energy resolution 
will systematically spill more photons to higher reconstructed energies than 
photons to lower reconstructed energies by sheer weight of numbers. This is a 
well known problem in spectral reconstruction of VHE sources. It is possible, 
with appropriate Monte Carlo modelling or bootstrapping, to compensate for this 
systematic trend if necessary.

\subsection{Sensitivity to burst intensity and asymmetry}
\label{sec:asymmetry}
\begin{figure}[htbp]
 \begin{center}
  \includegraphics[width=\textwidth]{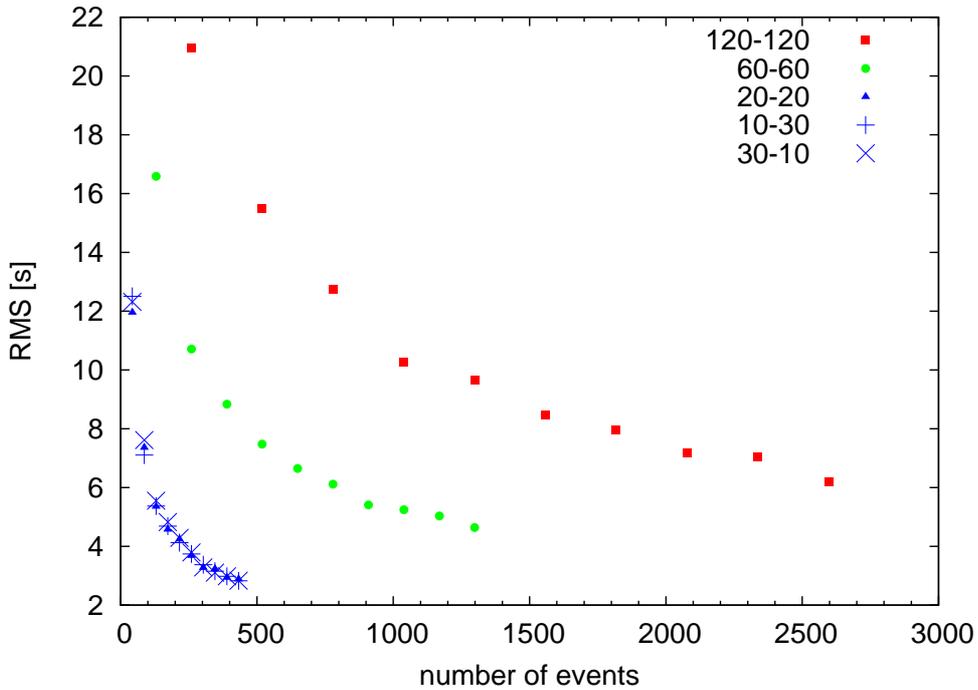}
  \caption {Sensitivity of the method in relation to the width of the
            burst and the number of events in it. The labels in the key 
            define respectively the rise and fall times of the profile. The 
            results are from MC simulations of 10,000 bursts with a maximum 
            count rate between 1-10Hz.}
  \label{fig:NumberEvents}
 \end{center}
\end{figure}

The burst intensity is another factor that will affect the sensitivity of the 
algorithm, since it will limit the photon statistics available to construct the 
CDFs. This is shown in figure~\ref{fig:NumberEvents} and was tested over a 
similarly-generated set of Gaussian profiles as before. Here bursts were 
generated with differing numbers of events, between 50-3000, accounting for 
different count rates (corresponding to $I_{max}$ between 1-10Hz) and for 3 
different burst widths, with rise/decay times varying between 10-120s. 

For a given burst width, the effect of increasing the number of events in the 
light curve is to reduce the RMS of the recovered dispersion parameter. From a 
certain number of events onwards, and depending on the width of the burst, the 
distribution tends toward a plateau and little improvement in the RMS is 
obtained by further increasing the number of events. As noted earlier, the 
sharper the burst, the earlier this plateau is reached. Finally, we have also 
tested for any effects due to profile asymmetry by maintaining the total burst 
width and varying the ratio of rise/decay time of the flare. The results plotted 
in Figure~\ref{fig:NumberEvents} show that the method is not affected by any 
intrinsic burst asymmetry, but only by its overall width.

\subsection{Sensitivity to energy spectrum}
\label{sec:SpectralIndex}
\begin{figure}[htbp]
 \begin{center}
  \includegraphics[width=\textwidth]{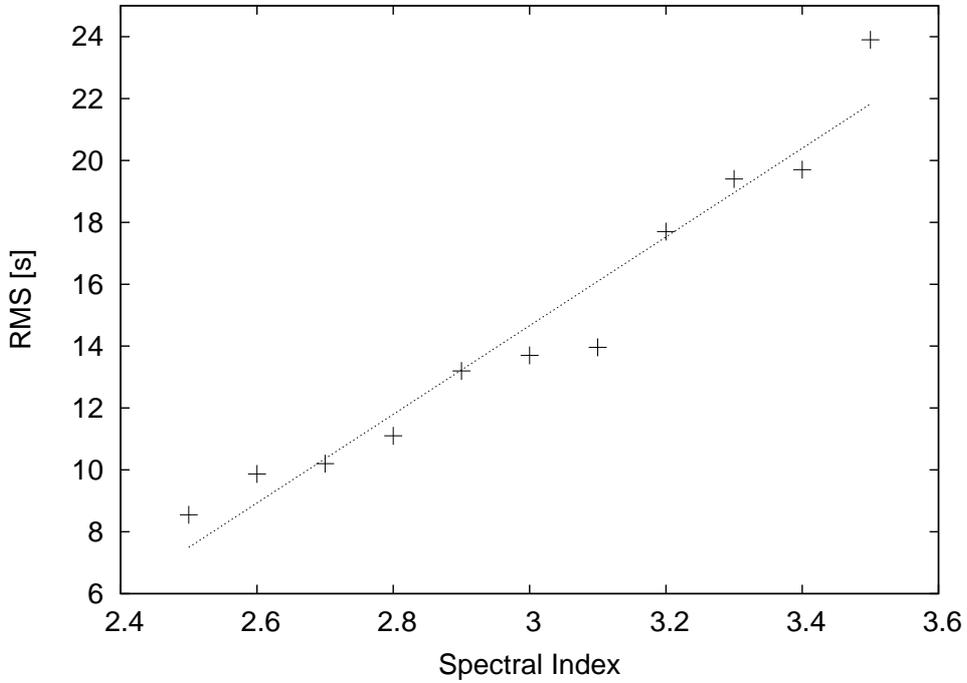}
  \caption {The effect of the spectral index on the uncertainty in the 
            recovered dispersion. The simulated bursts have a width of 120 s, 
            meaning that the recovered dispersion's RMS varies from 0.5-20\% 
            of the burst width, for an index $\Gamma$ going from -2.5 to -3.5.}
  \label{fig:SpectralIndex}
 \end{center}
\end{figure}

The observation of a single burst or flare is not going to provide definitive 
evidence (or refutation) of energy dependent dispersion due to QG. Instead, a 
number of sources demonstrating a consistent behaviour for a range of redshifts 
will be necessary to be able to confidently determine if such an important 
effect exists, and to disentangle it from source-intrinsic lags. Even if the 
intrinsic spectrum for a given source type is identical between objects, the 
interaction of the gamma-rays on the diffuse extra-galactic background radiation 
will lead to a softening of the observed spectral index with increasing 
redshift. The number of very high energy events is intimately related to the 
energy spectrum; to quantify the systematic uncertainty introduced by this 
effect in the estimate of $\tau^{*}$, we simulated 10,000 Gaussian shaped 
lightcurves, with 120\,s rise and fall times and a maximum count rate of 3Hz, 
for a range of indices $\Gamma$ between -2.5 and -3.5. The RMS of the recovered 
$\tau^{*}$ is plotted in figure~\ref{fig:SpectralIndex}, demonstrating an 
approximately linear deteoriation on the determination of $\tau^{*}$ as the 
spectrum softens. This effect is also easy to understand as the softening of 
the spectrum represents a depletion of high-energy photons from the high-energy 
CDF.

\subsection{Burst analysis window}
\label{sec:window}
Until now we have treated simulated isolated bursts, for which we are confident 
it is straightforward to find a reasonable signal-to-noise ratio above 
background from which we can define the burst to start/end. When analysing 
transient events within a real light-curve, as will be done in the next section,
it is important to consider the effects of confusion and under-sampling of 
individual bursts. If the burst is adjacent to other structures within a complex
lightcurve it might be difficult to define with precision its start and end 
times, and a superposition of different features might be unavoidable. In 
particular, the highest energy, most-lagged events could then fall outside an 
inappropriately chosen analysis window, thus affecting the profile 
reconstruction. Also, if the burst is at the edge of an observation run, data 
could be missing for part of the flare, this loss of information will also be 
energy-dependent if there are lags in the light-curve and this is likely to 
affect the performance of the reconstruction.

To test for these effects and assess if a proper reconstruction of the original 
profiles of lagged light-curves is still possible within our framework, we 
performed two sets of MC simulations, for which we generated two groups of 
10,000 Gaussian bursts with 500 events each, a spectral index of $\Gamma = -2.5$
and an energy resolution of 20\% was used. 

\begin{figure}[htbp]
 \begin{center}
  \includegraphics[width=\textwidth]{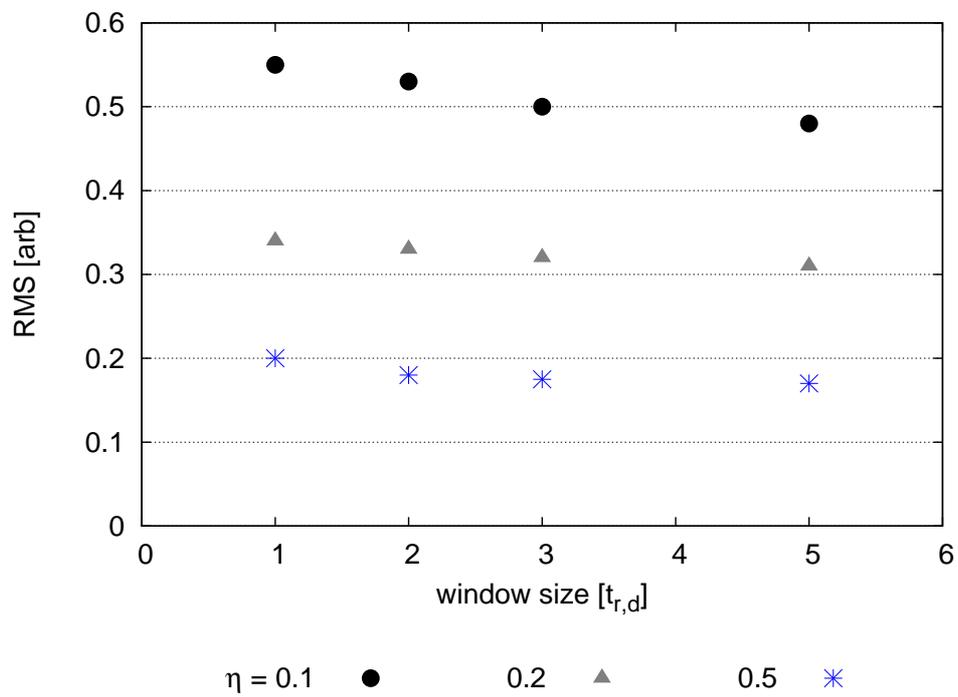}
  \caption {Sensitivity of the Kolmogorov distance method in relation to 
    the size of a ``transparent'' window used to reconstruct the CDFs 
    from the burst profile. The labels in the key define different data sets
    with different sensitivity factors $\eta =$ 0.5 (star), 0.2 (triangle), 0.1
    (circle). The results are from MC simulations of 10,000 bursts generated 
    from a Gaussian shape and an associated energy resolution for each event of 
    $\sim 20\%$.}
  \label{fig:TransparentWindow}
 \end{center}
\end{figure}

For the first set, represented in 
figure~\ref{fig:TransparentWindow}, the analysis considered a series of windows 
around the peak position of the burst of widths equal to 1, 2, 3 and 5 times 
the combined rise/fall time ($t_{r,d}$) of the burst (where these relate to the
time for full-width-half-maximum $t_{r}+t_{d}=t_{FWHM}$), to simulate different 
degrees of under-sampling. In 
this case, a so-called ``transparent window'' was used. This means that the CDFs
are built only with the events that fall within the time window boundaries after
the dispersion cancellation has been applied, but for each different value of 
$\tau$ events are allowed to pass into and out of the window's boundaries -- 
updating the CDF at each new step of the algorithm. 

The result is that a very narrow window around the burst affects the accuracy of
the reconstruction, increasing the RMS by up to 20\% when only the FWHM around 
the burst peak is used to build the CDF. This degrading effect can be understood
as a consequence of an ill-defined shape for the CDF.
The effect is present for all the 
range of sensitivity factors tested, being more pronounced for smaller $\eta$. 
This suggests that in using the method one should attempt to include as much of 
the burst as possible into the analysis, in order to include the most possible 
information on the profile shape for the CDF comparisons. An arbitrary choice 
of a narrow window about the peak of the burst to artificially reduce $\eta$ 
does not improve the results, due to a loss of information in the CDFs on the 
shape of the wings of the distribution. Of course this observation is no 
prescription for the analysis. Ideally the time window around the burst should 
be at least 3$t_{FWHM}$, but in the case that the burst is confused with other 
features the analyser should simply be aware of this degrading factor when 
determining the confidence intervals for $\tau$.

This also demonstrates that whilst individual sub-flare features may be
resolved, they can still potentially influence each other. If a train of 
bursts is too closely aligned with respect to each other for an individual 
analysis to be conducted, then our simulations show that the critical factor is 
to account for the rise/fall parts of the profile of the first/last burst 
respectively. In such cases, the best approach to the analysis is to consider 
the train of bursts together, rather than trying to split the bursts into
ill-sampled individual features.

As a final note, in the same way that events pertaining to the burst can be 
selected out of the analysis window, events not pertaining to the burst can also
contaminate the analysis during the cancellation procedure, for instance from 
background events or from superposing bursts closely aligned in time. This will
produce similar kinds of effects as the case treated in 
figure~\ref{fig:TransparentWindow}. As before, a compromise has to be found 
(if necessary via dedicated simulations with a configuration as similar as 
possible to that of the real lightcurve) between reducing the contamination and 
sampling well the profile of each individual burst.

\begin{figure}[htbp]
 \begin{center}
  \includegraphics[width=\textwidth]{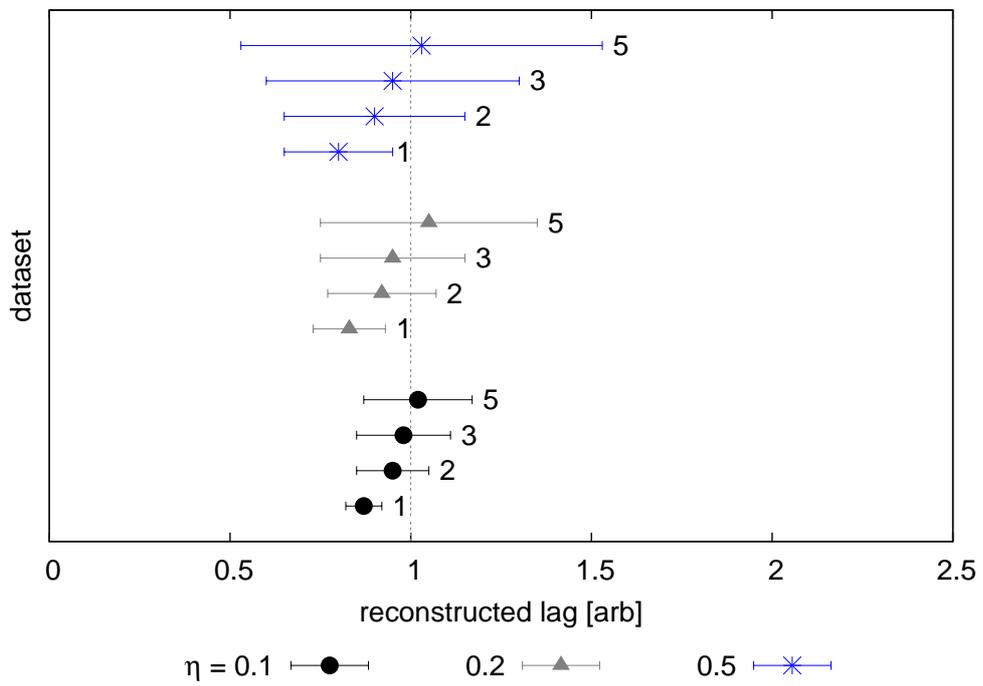}
  \caption {Sensitivity of the Kolmogorov distance method in relation to 
            the size of the ``opaque'' window used to construct the CDFs from 
            the burst profile. The different datasets correspond 
            to $\eta =$ 0.1 (circles), 0.2 (triangles), 0.5 (stars)
            respectively. The number 1-5 within each dataset are for windows
            of 1-5 ($t_{FWHM}$) widths, respectively. The results are from MC 
            simulations of 10,000 bursts generated from a Gaussian shape, and 
            an associated energy resolution to each event of $\sim 20\%$. The 
            actual value of the induced lag is indicated by the dotted vertical 
            line.}
  \label{fig:OpaqueWindow}
 \end{center}
\end{figure}

There is a second class of situations when a burst will be under-sampled at the 
detection level rather than in the analysis procedure, for instance when the 
burst occurs close to the start/end times of an observation run. In this case, 
events are lost permanently. To simulate this effect we introduce a so-called 
``opaque window'' in the analysis, where events that are initially out of the 
analysis window for a $\tau=0$ will not be admitted in as $\tau$ changes during 
the steps of the cancellation algorithm. The results are shown in 
Figure~\ref{fig:OpaqueWindow}. In this case, the fact that we 
\textit{permanently} lose a high proportion of high energy events 
(because they are more spread or dispersed then the low energy ones) means that 
not only will the RMS be worsened, we preferentially lose the events that will
most accurately recover the correct dispersion.

The three different data groups represented in Figure~\ref{fig:OpaqueWindow} 
are for sensitivity factors $\eta$ equal to 0.5, 0.2 and 0.1. Note that the 
case of longer duration bursts is the most affected, simply because in this 
case more high-energy events are permanently lost from the burst window, 
relative to the low-$\eta$ case. Within each dataset, points 1-5 indicate the 
size of the window in units of $t_{FWHM}$. Here, the permanent loss of 
information about the most-lagged events mean that the true lag $\tau^*$ is 
reconstructed wrongly the closer the burst maximum is to the start/end of the
observation window. Therefore, as before, the conclusion is that windows as 
wide as $\sim 3t_{FWHM}$ around the flare peak give the best compromise between 
burst width and information content. Another conclusion from this analysis is 
that the search for lags in a flare for which a large portion is missing from 
the burst (more than 1/4 of the the total number of events) is certainly not 
recommended, so observation strategies that maximise the amount of on-source
time in lightcurve monitoring (such as \cite{OrbitMode}) are definitely 
encouraged. 

In any case, as before, a complete simulation, or bootstrap, of the observed 
light-curve is the best way to assess the correct RMS from the combined effect 
of all cases discussed in this section. An example of this approach to the 
analysis of real data-sets is given next.

\section{Application to PKS\,2155-304 flare simulations}
\label{sec:PKS2155}
In the previous section we have discussed all the principal factors contributing
to the uncertainty in the reconstructed lag. We have illustrated those by the 
simple case of isolated Gaussian bursts. We are now in a position to test the 
efficacy of the method to recover a dispersion in some realistic lightcurves. 
To do so, we move from the simple tests on Gaussian profiles to work on 
simulated datasets based on fitted profiles for the large flare of PKS 2155-304 
from 2006 \cite{bigflare}. For consistency with previous work done on the 
PKS\,2155 data set, and to enable people to reproduce our work, we use exactly 
the same profile fits presented in the original H.E.S.S. publication, instead 
of searching for and separating the individual bursts ourselves 
(step 1 of the analysis procedure). This introduces a binned and parameterised 
aspect to the method, for that is how the lightcurve features were identified in
\cite{bigflare}. To keep the method truly non-parametric and unbinned, an
approach such as the Bayesian Blocks \cite{BayesianBlocks} algorithm to search 
for the time window cuts could be applied and is recommended, but it does not 
change the results we present here as this exceptional flare of PKS2155-304 is 
well resolved.

\begin{figure}[htbp]
 \begin{center}
  \includegraphics[width=\textwidth]{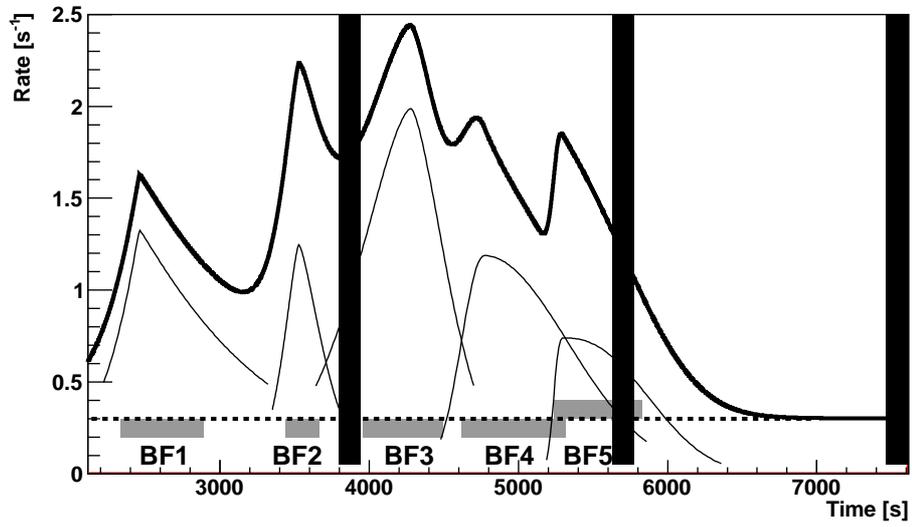}
  \caption {The parent population light curve of the MJD 53944 PKS 2155-304 big 
            flare event simulation. The individual bursts BF1-5 (see
            table\ref{tab:flares} are marked by the thin lines and the dashed 
            line shows the constant signal ($\sim 10\%$ of the maximum count 
            rate) onto which the bursts are superimposed. Values are 
            renormalised from the flux values in \cite{bigflare} to count rates 
            here. The heavy line denotes the cumulative light curve. The grey 
            shaded regions mark out the location and extent of the 1-$t_{r,d}$ 
            windows for the bursts, the black bands correspond to the data gaps 
            due to observing run transitions (see section~\ref{sec:window}). 
           }
  \label{fig:2155lc}
 \end{center}
\end{figure}

There are 5 prominent flaring events (labelled BF1-5) noted for this lightcurve,
reproduced in figure~\ref{fig:2155lc}, and the relevant parameters for
the generalised Gaussian fits are reproduced in table~\ref{tab:flares}. 
The simulated event times are generated by random draws from a distribution 
described by Equation~\ref{eq:GeneralisedGaussian}, each flare summed to give
the total lightcurve. To each event time, an energy value is then randomly 
attributed from a power law distribution, with $E_{\gamma} > 120$ GeV. As there
was no evidence for any spectral variability (at the $\Delta\Gamma\geq0.2$ level
\cite{bigflare}) during the 
night, for simplicity we adopted a simple power law distribution with a 
spectral index of $\Gamma=-3.5$ in our simulations; changing the model to the 
broken power-law fit given in \cite{bigflare} makes no difference to the general 
conclusions discussed here. The error in the energy reconstruction of a single 
event is dominated by systematic uncertainties and is estimated to be of the 
order of 15\% throughout the entire energy range. In reality the energy 
resolution is a function of energy, improving for higher energies, and so this 
value can be taken as a worst case scenario. Then, to simulate the energy 
dependent dispersion, a systematic delay $\tau$ was applied to each photon's 
true energy and the recovery procedure was carried out based on the 
instrumentally smeared energy values.

\begin{table}[htbp]
\begin{center}
\begin{tabular}{c|ccccc}
\hline
Flare & $t_{max}$ & Max. Rate & $\sigma_r$ & $\sigma_d$ & $\kappa$ \\
      & [s] & [Hz] & [s] & [s] & \\
\hline
BF1 & 2460 & 1.33 & 173 & 610 & 1.07 \\
BF2 & 3528 & 1.25 & 116 & 178 & 1.43 \\
BF3 & 4278 & 1.99 & 404 & 269 & 1.59 \\
BF4 & 4770 & 1.19 & 178 & 657 & 2.01 \\
BF5 & 5298 & 0.74 &  67 & 620 & 2.44 \\
\hline
\end{tabular}
\end{center}
\caption[]
        {Parameters used for the generalised Gaussian fit to the PKS\,2155-304 
         flare simulations, based on the original H.E.S.S. analysis results 
         \cite{bigflare}. 
         The third column (Max Rate) refers to the
         maximum count rate of each burst, corresponding to its peak
         flux at time $t_{max}$. The parameters $\sigma_{r}$ and
         $\sigma_{d}$ are the rise and decay time constants of each burst and
         $\kappa$ a measure of the sharpness of the peak (see text for details).}
\label{tab:flares}
\end{table}

\begin{figure}[htbp]
 \begin{center}
  \includegraphics[width=\textwidth]{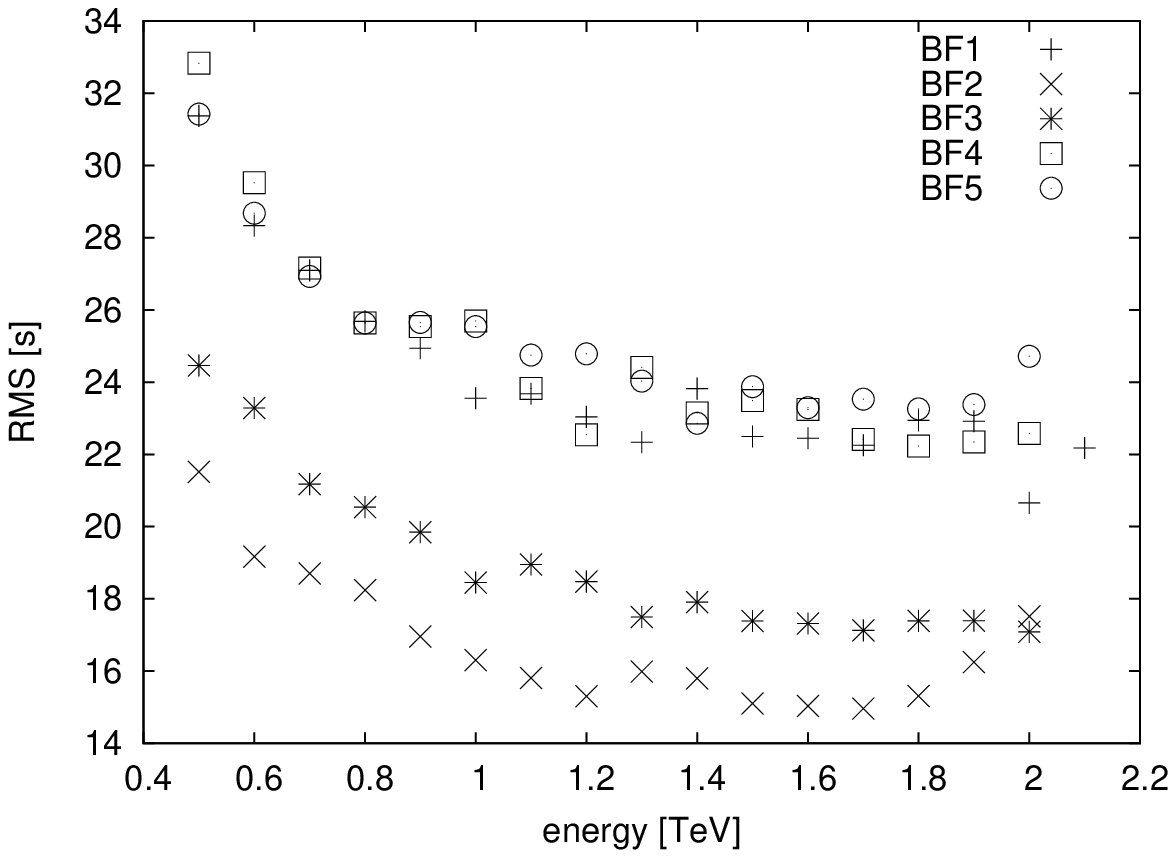}
  \caption {Effect of the choice of the energy cut for the high energy band
   on the accuracy of the determined dispersion measure based on Monte
   Carlo simulations of the burst profiles BF1-5 (see text for details).}
  \label{fig:EnergyCutBF}
 \end{center}
\end{figure}

\begin{figure}[htbp]
\begin{center}
\includegraphics[angle=270, width=0.3\textwidth]{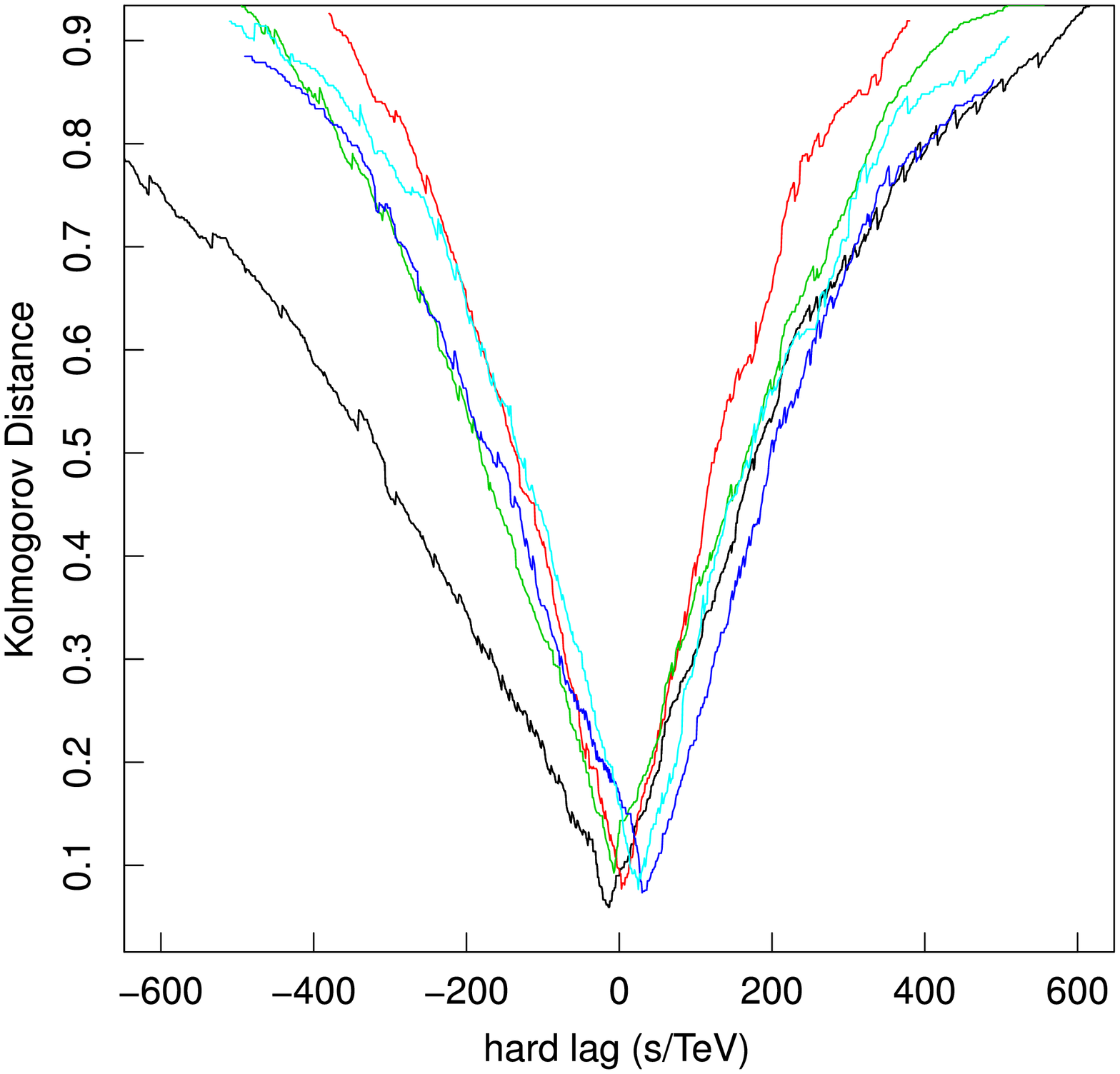} \\
\includegraphics[width=0.5\textwidth]{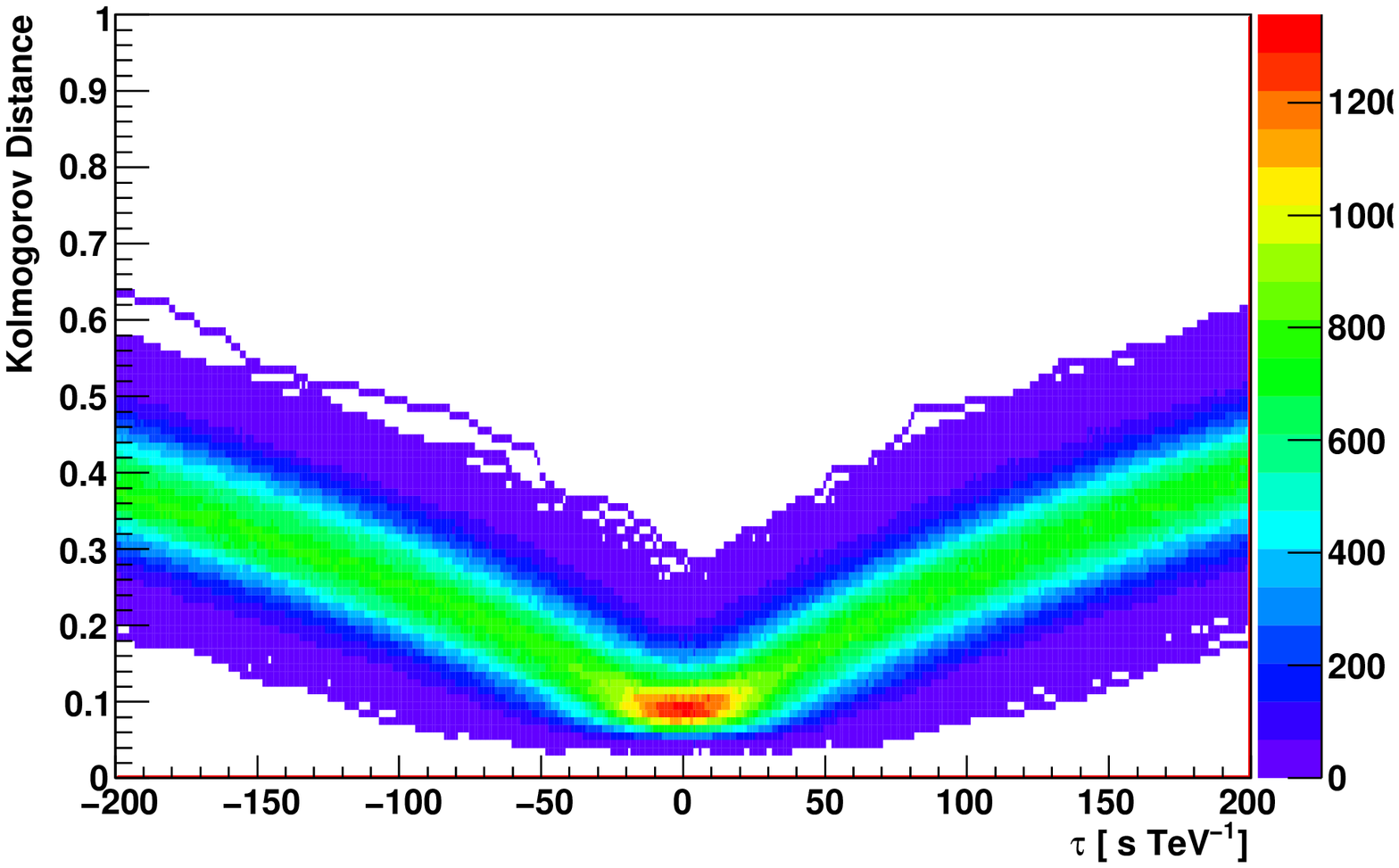} \\
\includegraphics[width=0.5\textwidth]{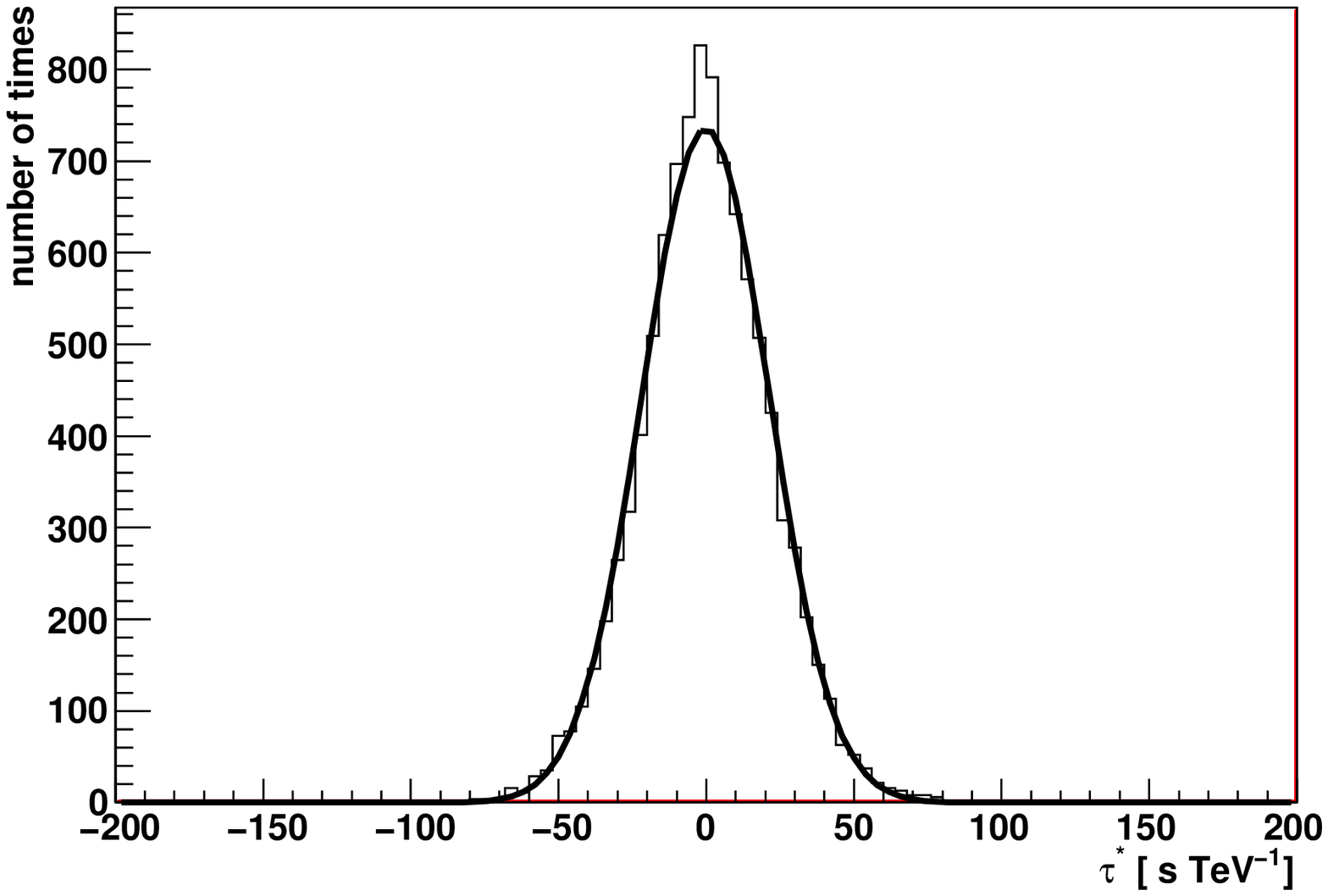} \\
\caption[]
        {{\bf Top panel:} Kolmogorov distance profiles for different values of $\tau$ 
         from the the search of energy-dependent dispersion in the bursts BF1-5 of PKS
         2155-304. The analyses were performed in steps of 1 s/TeV.
         {\bf Middle panel:} 2-D histogram of the Kolmogorov distance calculated 
         for each step of $\tau$ in 10000 simulated lightcurves for the 
         profile of BF2, with no dispersion introduced. A minimum is always 
         reached with some variance around 0\,s/TeV.
         {\bf Bottom panel:} Histogram of $\tau^{*}$ for each of the
         minimum $D_{k}$ values found in the simulations of BF2 shown in the middle panel, 
         centred on 0. The standard deviation of a Gaussian fit to this 
         distribution is used to estimate the limiting accuracy for 
         reconstruction of $\tau^{*}$.
        }
\label{fig:TestBF2}
\end{center}
\end{figure}

In a real-case analysis like the one shown here, there are two non-trivial steps
(numbers 1 and 2 of the list shown in section 3) that must be considered: (i) 
the choice of the analysis window around each burst and (ii) the choice of the 
energy boundaries to construct the CDFs. Figure~\ref{fig:EnergyCutBF} shows the 
results of our analysis on the effect of the choice of the high energy cut on 
the RMS of the reconstructed dispersion parameter for the individual flares. 
The improvement in the RMS as the cut moves away from the soft energy band is 
notable. It is followed by the presence of an optimal plateau around and above 
1\,TeV and worsening RMS above 2\,TeV due to a loss of event statistics. All 
this reproduces what was seen in the ideal flare shape case of 
section~\ref{sec:EnergyCuts} and shows the choice of 1\,TeV for 
$\mathrm{H_{min}}$ to be a good one. The uncertainty in the reconstructed lags 
(in s/TeV) were determined from Monte Carlo simulations performed for each 
individual burst. The top panel of figure~\ref{fig:TestBF2} shows an example of 
values of the Kolmogorov distance $D_{K}$ for each different dispersion 
parameter $\tau^{*}$ tested in the analysis. The middle panel shows the 
distribution of $D_{K}$ versus $\tau^{*}$ for 10,000 realisations of BF2, 
from which confidence intervals for the lag were derived, as shown in the 
lower panel histogram.

\subsection{Testing the recovery of a known induced dispersion}
\label{sec:known}
\begin{figure}[htbp]
 \begin{center}
  \includegraphics[width=\textwidth]{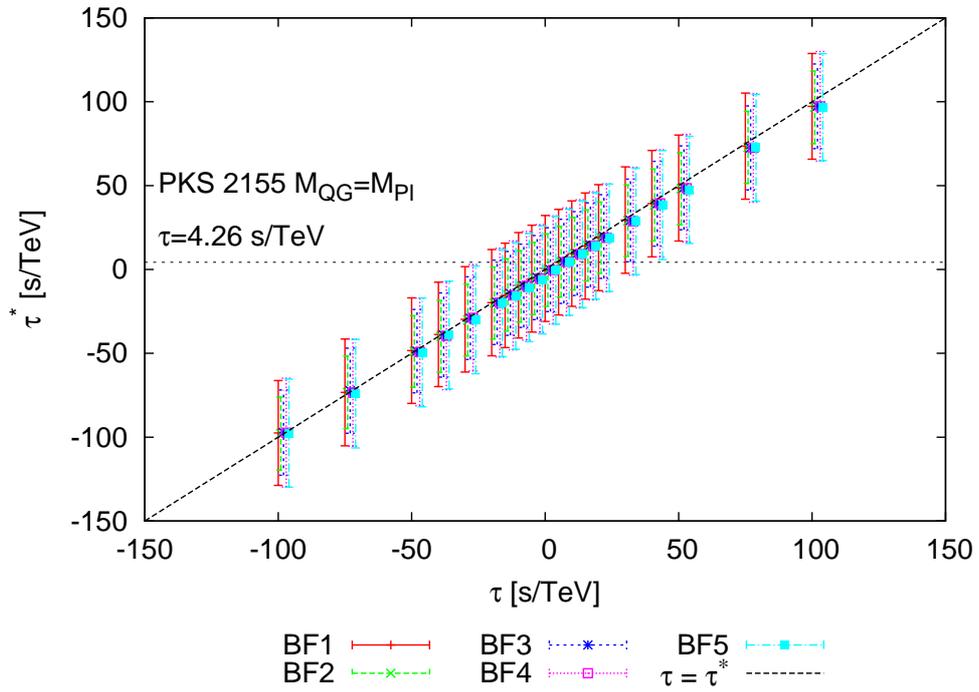}
  \caption {The accuracy to which the method can recover a fixed dispersion
  parameter introduced into the light-curves representing each of the 
  sub-flares of PKS\,2155-304. Each point is the average dispersion
  estimated from 10,000 simulated lightcurves and the errors represent
  the RMS of the recovered dispersion parameter, $\tau^{*}$. Values 
  corresponding to the different flares are slightly offset on the x-axis for 
  visual clarity.
  The dashed line is a guide to the eye of $y=x$, not a fit to the data. 
  The dot-dashed line represents the expected linear dispersion for a Planck 
  mass scale of quantum gravity from a source at the redshift of PKS\,2155-304.}
  \label{fig:RecoverDispersion}
 \end{center}
\end{figure}

To demonstrate the efficacy for recovering a dispersion, for each simulated 
lightcurve in figure~\ref{fig:RecoverDispersion} we introduced an artificial 
dispersion between $-100\leq\tau\leq100$ s/TeV, which we aimed to recover with 
the dispersion cancellation algorithm and minimisation of the Kolmogorov metric. 
The algorithm was applied to each of the five major burst features in the 
dataset, BF 1-5, generating five sets of independent measurements. Whilst the 
RMS of the recovered dispersion $\tau^{*}$ leaves uncertainty in the true 
dispersion $\tau$, the mean of the recovered dispersion is an accurate
reflection of the true dispersion. This shows that the measurement of multiple 
flares from a single object could be used in combination to give a more accurate
estimate of any induced dispersion, since they should be the same for all flares
if they have an origin in propagation effects. Also, the accuracy of the 
recovery over a large range of the parameter space demonstrates that a number of
objects at different redshifts could then accurately trace a distance dependent 
(i.e. propagation induced) dispersion.

\subsection{Estimating the limits to QG measurements from the Kolmogorov metric}
\label{sec:limits}
It is known from previous studies of this dataset that there is no spectral 
dispersion present in the PKS 2155-304 big flare night lightcurve that is 
large enough to be detectable within current instrument's sensitivities. We 
therefore limit ourselves to using the simulations of the dataset as a test-bed 
for assessing the sensitivity of the method in a real-life case. The 
$E_\mathrm{QG}$ sensitivity limits that could be placed from the 
application of the algorithm to the big flares from PKS\,2155-304 are presented 
in table~\ref{tab:realflares}. We adopted a linear relation between the lag and 
energy of the photon ($\alpha=1$ in equation~\ref{eq:cancel}) for the $E_{QG}$ 
limit estimates in the final column. The mean energy difference 
$\langle\Delta E\rangle$ between the low- and high-energy profiles is also
given. Note that the sensitivity limits obtained here are comparable with the 
most constraining results achieved to date from more complex analyses of this
particular flare \cite{likelihood2155}. This result shows that this simple 
method has the potential to probe QG effects to {\it at least} the same levels 
achieved by the current tests using AGN, with the advantage that, by separating 
the light-curve into multiple bursts, from a single dataset we can derive 
multiple independent measures of the same quantity.

\begin{table}[htbp]
\begin{center}
\begin{tabular}{c|ccccc}
\hline
Flare & $\langle\Delta E\rangle$ & $\sigma_{\tau}$ & $E_{QG}$\\
      &     [TeV]    & [s/TeV]    & [$10^{19}$\,GeV] \\
\hline
BF1 & 1.42 & $\pm$24.1 & $>0.15$\\
BF2 & 1.23 & $\pm$16.4 & $>0.19$\\
BF3 & 1.40 & $\pm$18.6 & $>0.2$\\
BF4 & 1.25 & $\pm$24.4 & $>0.13$\\
BF5 & 1.29 & $\pm$24.5 & $>0.14$\\
\hline
\end{tabular}
\end{center}
\caption[]
        {Results for each of the PKS2155 big flare night sub-flares  
         $\langle\Delta E\rangle$ is the mean energy difference between the 
         low-and high-energy CDF. $\sigma_{tau}$ corresponds to the standard 
         deviation of a Gaussian fit to the histogram of $\tau^{*}$ from 
         10,000 simulated lightcurves. The sensitivity limits on $E_{\rm{QG}}$ 
         correspond to the 95\% confidence levels from the uncertainty in the 
         recovered $\tau^{*}$. Note that these values are not real 
         estimates of $E_{\rm{QG}}$ limits from the PKS 2155-304 observations,
         but sensitivity estimates to searches in a simulated dataset.} 
\label{tab:realflares}
\end{table}

It is apparent in figure~\ref{fig:RecoverDispersion} that any dispersion 
$|\tau| < 50$\,s/TeV is unlikely to be confidently determined from an 
individual flare observed with the current generation of Cherenkov telescopes. 
The exceptional flares described here each have too small a value of $\eta$ 
to resolve such a signature individually, the even longer typical widths of 
observed VHE AGN flares only serves to reinforce this point. Nevertheless, the 
fact that lags can be accurately reconstructed over a large range of $\tau$, 
demonstrated by the correct mean value of the reconstructed lags, supports the 
idea that the combined results of a number of flares from an individual source 
could improve the $E_{\rm{QG}}$ limits considerably for each given object.

\section{Discussion and Conclusions}
\label{sec:con}
We presented here an unbinned, non-parametric method of testing for energy 
dependent dispersion in a lightcurve that is sufficiently robust to work under 
the constraints of scarce counting statistics and the modest energy resolution 
expected for VHE gamma-ray data analysis, the latter issue having been noted as
a limiting factor in other simple dispersion cancellation methods, eg 
\cite{scargle08}. The applicability of the method is based on the fact that a 
random distribution of events in energy will give rise to indistinguishable 
time profiles (or CDFs) that can thus be directly compared using some kind of 
statistical metric, among which the {\it Kolmogorov metric} was found by us to 
be best suited to our purposes. When used in conjunction with unbinned 
algorithms for identifying variability features like flares in lightcurves, 
such as for example the Bayesian Blocks algorithm\cite{BayesianBlocks}, the 
analysis chain would be an entirely non-parametric way to search for energy 
dependent dispersion (though such an analysis is beyond the scope of this 
rather specific discussion).

We have discussed in detail the factors contributing to the uncertainty in 
the determination of a lag, and the method was subsequently applied to the 
challenging case study of looking for a small magnitude dispersion 
expected from a specific QG-induced LIV, following the form of
equation~\ref{eq:perturbation}. It is already known that there is no 
dispersion present in the PKS\,2155-304 big flare night lightcurve that is any 
larger than the limits achievable by our method, presented in 
table~\ref{tab:realflares} \cite{xcol2155}, therefore we did not attempt to 
derive further limits using that exemplary dataset. It should be noted, 
however, that the limits for each of the simulated sub-flares are at least a 
factor of two better than the cross-correlation method of \cite{xcol2155} and 
comparable to the much more complex analysis of \cite{likelihood2155}; but 
while those two methods used the entire night's lightcurve as a single dataset 
to derive their limits this method has the advantage of being able to treat 
distinct bursts within the lightcurve as independent tests. 
Analysing a larger number of individual sub-flare features with the method 
presented here, and posteriorly combining a statistical sample of flares to 
obtain a single estimate (or limit) on dispersion could correspondingly 
make for yet more constraining limits on LIV.

In section~\ref{sec:SpectralIndex} we found the intuitive result that the 
uncertainty in any measured dispersion scales inversely with the hardness of 
the energy spectrum. This also leads to the slightly counter-intuitive 
consequence that a nearer AGN such as Mrk\,421 in a high state (eg 
\cite{Blazejowski2005, Mrk4212011}) could provide as constraining a limit, if 
not more so, than a more distant object (like PKS\,2155-304) if it showed 
similar variability timescales. This is solely due to the harder observed 
spectra, meaning an increased number of high energy photons being detectable 
from the nearer object (due to less absorption by the extragalactic background 
light). This could give a smaller uncertainty on the dispersion even if the 
expected magnitude of the delays are smaller. At a redshift of $z\sim0.03$ for 
Mrk\,421, the expected dispersion would be $\tau \sim 1$\,s/TeV for Planck mass 
scale quantum gravity; with a next generation observatory such as CTA \cite{CTA} 
the number of photons above 10\,TeV in minute scale flares could potentially 
push the expected sensitivity $\eta$ an order of magnitude beyond what is 
achievable today with AGN studies. This means that such AGN studies could 
perhaps surpass even the current Fermi limits on GRB~090510 \cite{FermiLimits} 
making for a useful independent confirmation of that result. The combined 
information from different source classes and over a sufficient range in 
redshift would then enable to distinguish between intrinsic source and external 
propagation induced dispersion effects.

The performance on the method has been tested on a specific case, but this 
method is, of course, not just limited to searches for a linear expansion
term and VHE gamma-ray observations provide the highest energy photon datasets 
for probing the quadratic term of equation~\ref{eq:perturbation} (e.g., 
\cite{likelihood2155}). The simple isotropic dispersion scheme of 
equation~\ref{eq:perturbation} also neglects a number of aspects of potential 
Lorentz violation effects, such as birefringence and direction dependent vacuum
co-efficients for Lorentz invariance (see eg equation 145 of 
\cite{Kostelecky09}) which could make up interesting follow-up studies with
either this or next generation.gamma-ray telescope datasets. Neither is our 
method just limited to ground based VHE gamma-ray datasets, the extended 
baseline for the Fermi GRB datasets can more than compensate for the lower 
photon energies (with regard to the linear term, \cite{FermiLimits}), and tests 
for birefringency effects (which instead cause a spread in the wave packets 
\cite{Kostelecky09}) have been made using INTEGRAL observations of the highly
polarised GRB\,041219A. Unfortunately it is not currently possible to determine
the polarisation of gamma-rays from ground based obervations, so this case has
not be dealt with here, but broadening of the lightcurve can also arise in
certain models that predict a stochastic, super-luminal aspect to the refractive
index (eg \cite{superluminal}), though such tests may be better aided with a 
metric that is optimal in testing the change in kurtosis of the lightcurve, 
rather than the skewness, as used here. 

Most importantly of all, the method for testing for the presence of a 
dispersion is not limited to just testing for Lorentz invariance violation 
effects. Indeed any physical model that introduces a width or a delay in the 
emission, such as acceleration within the emission region (eg 
\cite{Bednarek08}) or due to cascading on the intergalactic magnetic 
field (eg \cite{Bfields}), can be adopted as the model for the time delay 
correction. These are all tests to be pursued in future work and not in the 
scope of this paper, which is to present the method, its benefits and 
limitations.

\section*{Acknowledgements}
We thank P.~M.~Chadwick for useful comments on the original work. We also thank 
the anonymous referees for many critical comments which helped to improve very 
much this manuscript. UBdA acknowledges a PhD Scholarship from the CAPES 
Foundation, Ministry of Education of Brazil. 


\appendix

\section{Light Curve Representation}
\label{app:lightcurve}
In order to apply the algorithm, it is necessary to define how to construct the 
CDFs from the low and high-energy event sequences of the burst under study 
(step 3 of the analysis procedure). Given that the Kolmogorov distance is a 
metric for probability distributions, the event sequence must be normalised. 
Since the dataset is composed of time- and energy-tagged events, the 
cancellation will be applied to every photon individually so that none of the 
available information is left out of the analysis. The simplest choice for 
representing the data is therefore to construct empirical CDFs for both the 
low- and high-energy profiles as step functions from the original event 
sequence, according to the following rule:
\begin{equation}
\label{eq:cdf1}
 CDF:~ F(t_{i}) = i/N,
\end{equation} 
where $t_{i}$ is the time of the i$^{th}$ event in the sequence, and $N$ is the 
total number of events in the sequence. In this construction, the height of each
step is constant and equal to $N^{-1}$ (the CDF is defined between 0 and 1), and
the length of each step equals the waiting time between events in the sequence. 
All the timing information of the temporal sequence is thus explicitly 
preserved in this representation.
 
\begin{figure}[htbp]
  \centering
  \includegraphics[angle=270, width=\textwidth]{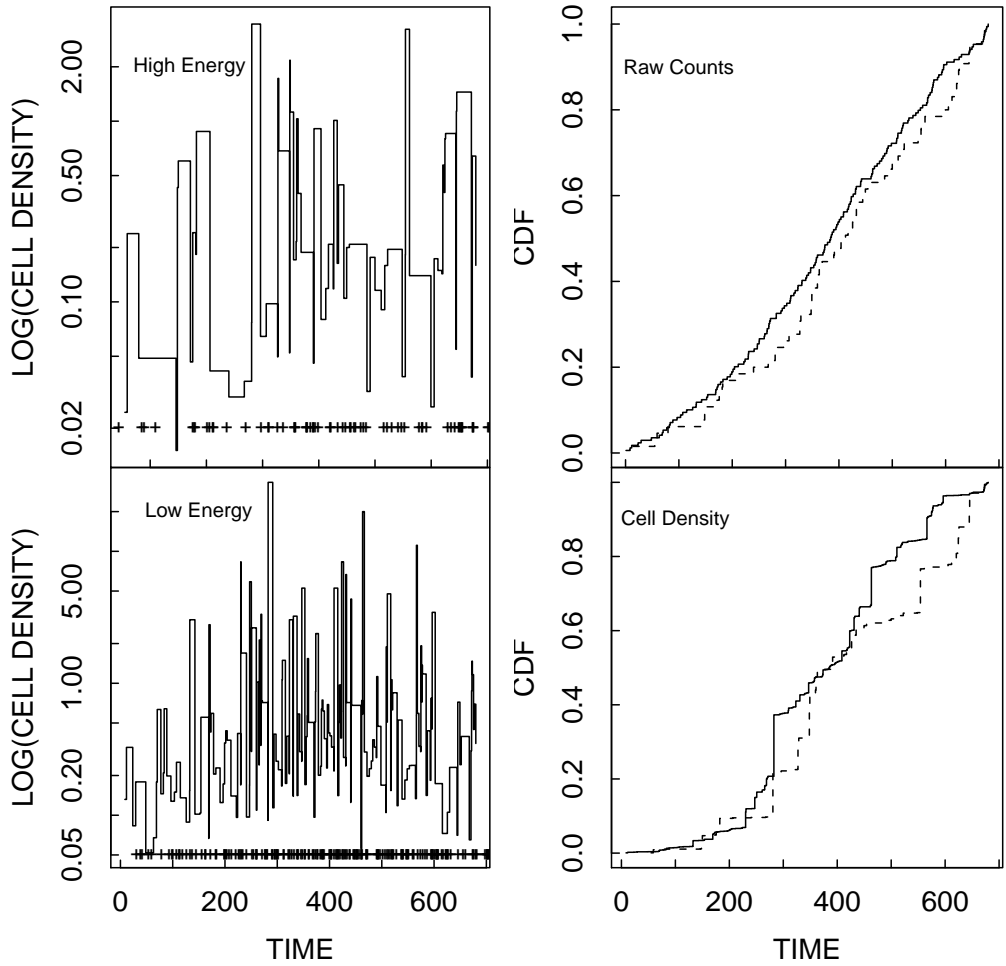}
  \caption {Choice of the light-curve representation. The panels on the
    left show the cell density representation for the low- and
    high-energy components of flare BF1 of PKS 2155-304 (see
    \ref{sec:PKS2155} for nomenclature of the flares). The right
    panels show the correspondent CDFs for the two light-curve representations 
    discussed. Note that the {\em raw events} representation shows
    considerably less ``raggedness'' in the CDFs than the {\em cell density} one, 
    and is therefore more appropriate for calculating the Kolmogorov distance cost-function.}
  \label{fig:Kolmogorov}
\end{figure}
 
A different representation of the dataset was proposed in 
Scargle et al. (2008)~\cite{scargle08}, and can be used as an alternative way 
of constructing the CDF. In this representation, the dataset is tesselated so 
that the photon sequence is represented by a series of cells of width $dt_{i}$ 
constructed around each event $i$. A cell density is then defined by the rule 
$x_{i} = 1/dt_{i}$, which can be interpreted as the instantaneous rate of the 
process at time $t_{i}$, which is later normalised into a discrete probability 
distribution: $p_{i} = x_{i}/\sum x_{i}$. The CDF in this case would be:
\begin{equation}
\label{eq:cdf2}
 CDF:~ F(t_{i}) = \sum_{t~<~t_{i}} p_{i},
\end{equation}

For the application of the Kolmogorov distance metric, we found that the first 
representation in equation~\ref{eq:cdf1} is more appropriate. This is because 
the magnitude of the cell densities representation can be dominated by spikes 
resulting from very small inter-event times in some cells, originating from the 
noise of the Poisson process, which will introduce excessive ``raggedness'' in 
the CDF representation. This can be seen in the right panel of 
figure~\ref{fig:Kolmogorov} where we compare the low- and high-energy CDFs from 
a simulated burst profile based on the burst BF1 of the VHE flare of 
PKS 2155-304 observed with H.E.S.S. In this case, both profiles superpose, as 
there is no significant dispersion present, but the cell density representation 
results in additional fluctuations in the constructed CDFs. A way to circumvent 
this problem within the cell representation is to adopt a logarithmic scale for 
the density -- for example $x_{i}=\rm{log}(1/dt_{i})$ -- which better recovers 
the shape of the profile.



\bibliographystyle{elsarticle-num}
\bibliography{dispersion-measure.bib}







\end{document}